The effect of periodic spatial perturbations on the emission rates of quantum dots near graphene platforms

X. Miao[1], D. J. Gosztola[2], X. Ma[2], D. Czaplewski[2], L. Stan[2] and H. Grebel[1]*

[1] Electronic Imaging Center and ECE Dept., New Jersey Institute of technology (NJIT), Newark, NJ 07102, USA. grebel@njit.edu
[2] Center for Nanoscale Materials, Nanoscience and Technology Division, Argonne National Laboratory Argonne, IL 60439

**Abstract:**

Quenching of fluorescence (FL) at the vicinity of conductive surfaces, and in particular, near a 2-D graphene layer has become an important biochemical sensing tool. The quenching is attributed to fast non-radiative energy transfer between a chromophore and the lossy conductor. Increased emission rate is also observed when the chromophore is coupled to a resonator. Here we combine the two effects in order to control the emission lifetime of the chromophore. In our case, the resonator was defined by an array of nano-holes in the oxide substrate underneath a graphene surface guide. We demonstrated an emission rate change by more than 50% as the sample was azimuthally rotated with respect to the polarization of the excitation laser. Such control over the emission life-time could be used to control resonance energy transfer (RET) between two chromophores.

**Introduction:**

Quenching of fluorescence in the vicinity of conductors is well documented [1-2]. The growing interest in graphene – a mono, or a few layers of graphite – has extended the study of fluorescence-quenching to this unique film [3-10]. If the potential barrier between the graphene and the QD does not allow for a direct charge transfer, energy transfer may be advanced via dipole-dipole interaction [9-11]; such fluorescence resonance energy transfer, or FRET, is enabled through screening of the excited fluorophore by the free-carriers in the graphene film (a Förster process). For the energy transfer to be effective, the lifetime of the QD near the graphene needs to be shorter than the life-time of a stand-alone QD. The absorption of graphene (~2.3% per layer) ought to be compatible with the absorption of the CdSe/ZnS QD monolayer so that the film of dots will not screen itself out [12,13].

Intensity studies need to be complemented by time-resolved emission rates [13]. Concentration dependent signals [14], masking the conductor by relatively thick QD films [15] and charge coupling between nearby dots may obscure the local interaction with the conductor. Since the energy transfer depends on the distance between the graphene and chromophore, a spacer may control their mutual interaction. While very thin, this spacer - a 10-nm hafnia film on the graphene - in addition to the QDs and the graphene itself may construct a surface optical waveguide.

Coupling the excitation laser to this guide, or coupling the emission to free-space radiation modes may be conveniently made through a periodic array of nano-holes in the oxide substrate under the graphene layer. The array of holes also provides for spatial confinement of the surface mode, effectively increasing its propagation life-time but also increasing the emission rate of a nearby QD via an increase in the emission's density of states (DOS) (Purcell effect).

Screening near the Dirac point by charged carriers depends on the amount of charge placed within a small distance away from the graphene [16]. Again, local laser intensities, local chromophore concentration and other scattering may affect FL intensity variations. To a first-order, life-time parameters are not affected by the laser intensity but are affected by the local DOS [17-20]. The DOS for a 3-D system is proportional to the square of the radiation frequency, $\omega^2$. The DOS for a 2-D system is linearly proportional to $\omega$. Therefore, if energy is coupled to a 2-D guide before coupling to a 3-D free-space mode, then, the emission life-time may be prolonged.

Our ultimate goal is to explore energy transfer from one QD to another via a surface guide as a mediator. To do so effectively, we first need to control the lifetime of the energy donating QD through a simple mechanism; in our case, this will be an azimuthal rotation of the substrate. The process is broken into several steps (see also SI section):

1. Excitation of the chromophore (here, the QDs) by a pump laser at frequency $\omega_L$. The chromophore is relaxed and transfers energy at frequency $\omega_E$ to a 2-D graphene guide on an hole-patterned oxide substrate. The 10-nm hafnia on top of the graphene serves as a spacer between the graphene and QD and controls their mutual interaction. A surface mode is sustained due to the large refractive index of graphene ($n_{graphene}$~2.6), QDs and hafnia, but not necessarily through a plasmonic mode for which the dielectric constant of graphene needs to be negative. The latter effect is observed at longer IR wavelengths [21] and could, in principle, be observed through down-conversion of visible QD's emission, or, through parametric oscillations.

2. The excited QD dipole is coupled non-radiatively to a charge dipole in the graphene via energy transfer [9-11] at the rate of $\Gamma_{i1\rightarrow f1}$ with i1 - the initial, excited state of QD and f1 - the final state, the excited dipole in the graphene. The final state, f1 may transfer its energy to another QD nearby or thermally relax. If the graphene is coupled to a resonator (the periodic spatial pattern), then the QD may relax at a rate of $\Gamma_{i1\rightarrow f2}$ with i1 - the initial, excited state of the QD and f2 - the final electromagnetic state within the surface resonator. That mode may propagate back and forth along the surface resonator and eventually be coupled to free space modes or back to the lossy graphene film. Coherence in our case is achieved when the surface mode is at resonance with the local periodic perturbations; the intensity of the mode stays mostly within the structure holes as we shall see below. A third interaction channel between the standing surface mode and the dipole generated in the graphene may be possible. Its mutual coupling may be sensitive to nonlinear photonic, or phononic effects [22] and could result in energy exchange. We will not dwell on such effect but a discussion is provided in the supplementary information section (SI). Overall, our measurements were carried for fluorescence intensity values that were linear with respect to the laser intensity.

3. The surface mode is coupled to free-space radiation modes and detected by a faraway detector.

Furthermore,

(a) when all the other parameters are kept the same, the emission rate of a chromophore coupled to a 2-D system is *smaller* than a chromophore coupled to a 3-D system;

(b) the conductive graphene *increases* the emission rate through non-radiative energy transfer process, which is enabled by charge screening;

(c) the effect of a resonating spatial perturbation is to further *increase* the emission rate of the chromophore due to an increase in the DOS near resonance [17]. The measured rate is $\Gamma_{ET} = \Gamma_{i1\rightarrow f1} + \Gamma_{i1\rightarrow f2}$ in the absence of other nonlinear processes (see SI section). The process efficiency is $E \sim \Gamma_{ET}/(\Gamma_{ET}+\Gamma_D)$, where $\Gamma_{ET}$ and $\Gamma_D$ are, respectively the non-radiative rate of energy transfer and the radiative decay rate of a stand-alone donor, increasing or decreasing of $\Gamma_{ET}$ provides an active control over the entire process;

(d) a photon travelling back and forth within a resonating structure (namely, the surface guide with periodic perturbations) forms a standing wave at resonance conditions. The resonance conditions result in enhanced intensity at some particular tilt and azimuthal rotation angles with respect to the nano-hole array [13,23,24].

Here, we show how to control the life-time of a chromophore embedded in a hole-array, or above it by azimuthal rotation of the sample with respect to the laser polarization.

**Theoretical Considerations:**

Following [23], once coupled to the surface guide, the mode propagates in the x-y plane with a wavevector, $\boldsymbol{\beta}_s$. A standing wave is formed if a Bragg condition is met: $|\boldsymbol{\beta}_s - \mathbf{G}| = \beta_s$; $\mathbf{G}$ is the reciprocal wave vector of the spatial square array of holes with a pitch $\Lambda$. The wavenumber of the surface mode may be written as, $\beta \sim k_0 n_{eff} = (2\pi/\lambda_0) n_{eff}$. Here $\lambda_0$ is the free-space emission wavelength and $n_{eff}$ is the effective refractive index of the surface mode (including the 10-nm hafnia, the QDs and their ligand coating).

An efficient coupling of the surface mode to and from free-space mode occurs if momentum is conserved: $\boldsymbol{\beta} = \mathbf{k_o}\sin\theta + q\mathbf{G}$ with q integer. At normal incidence, we may pick up the x and y coordinate along the square hole-array coordinates, $\beta\cos(\phi) = q_1 G_x$ and $\beta\sin(\phi) = G_y = q_2 G$ with $q_{1,2}$ – integers and for square array, $G_x = G_y = G$ (see SI section). At normal incidence, coupling to the surface waveguide and the establishment of resonance conditions may occur simultaneously with the same angle $\phi$ and subwavelength patterns; for example if the scattering happens along the x-direction, m=2q [25]. A simplified numerical model is described in the SI section.

The simulations indicate that the propagation along the x-direction in the surface guide may be polarized along either the y-direction (parallel to guide surface) or z-direction (perpendicular to the guide surface) for excitation and emission modes. For the emission wavelength, $\lambda_0 = 575$ nm, resonance occurs at normal direction, $\theta = 0^o$ with $\phi = 0$. For the incident wavelength and $\theta = 0^o$, resonance occurs at $\phi = 45^o$ (along the cell's diagonal) and the coupling to the surface guide is made with every other hole-plane q=1/2. When excited by an s-polarizations (polarization parallel to the surface guide) there is a non-zero z-component (perpendicular to the guide surface) mostly in the air pillars. This implies that excited QDs, situated in, or nearby holes, are spatially correlated.

***(a) Polarization and dephasing:*** At normal incidence, the incident laser beam is polarized parallel to the guide's surface. Simulations suggest that the pillar-interfaced, graphene surface guide supports TE modes for the excitation and emitted wavelengths ($E_y$, parallel to the guiding surface). At normal incidence and through momentum conservation, the *s*-polarized excitation mode is coupled to two counter-propagating TE guided modes. The de-polarization [26] through relaxation of the excited carriers, from the excited state to the bottom of the conduction band, is small because this non-radiative process is very short (~100 fs) compared to the emission life-time (~1 ns). Both the excitation and the emitted wavelength may be coupled via the hole-array [25]. If, $\beta_L$, $\beta_e$ are the surface wavevectors for the incident and emitted modes, then for a co-linear case, $\beta_L+\beta_e+\mathbf{G}=2\pi n_{eff}\cos(\phi)/\lambda_L+2\pi n_{eff}/\lambda_e-2\pi/\Lambda=0$ within 1% if we assume $n_{eff}=1.15$ (see below).. For a square array we expect the transition rate to have a $90^o$ symmetry, considering two orthogonal Bragg reflectors along the x- and y-directions. Discussion on the periodical fit is provided in the SI section.

**Results and Discussions:**

In Table 1 we provide description of four samples that were prepared in various ways. Common to them is the average spacing between the QD and graphene (either from the above or below it). Scanning Electron Microscope picture of a bare patterned substrate is shown in Fig. 2a. Detailed description of the samples is provided in Fig. 2b,c and Table 1. Typical Raman spectra taken when the dots were deposited on top of the hafnia/graphene layer, or deposited under the graphene (while still with the hafnia on top) are shown in Fig. 1d.e. Raman maps of the 2D line for the two cases are shown in Figs 1f,g respectively and allude to the monolayer nature of the graphene. The maps are overlaid on an image of the substrate; some cracks in the hafnia are noted.

| Sample | QD deposition method | Placement of QD | Concentration | Spacer/top coat |
|---|---|---|---|---|
| S2 | spin | in holes | High | no/pmma |
| S7 | dip | on spacer | High | yes/no |
| S8 | spin | in holes | Low | yes/no |
| S9 | spin | on spacer | Low | yes/no |

Table 1. Sample description. Spinning was made at 2500 RPM for 30 sec; QD concentration was: High=1 mg/mL; Low 0.25 mg/mL; dipping was made with high concentration at a speed of 2 mm per minute; spacer is the 10-nm hafnia on top of the graphene.

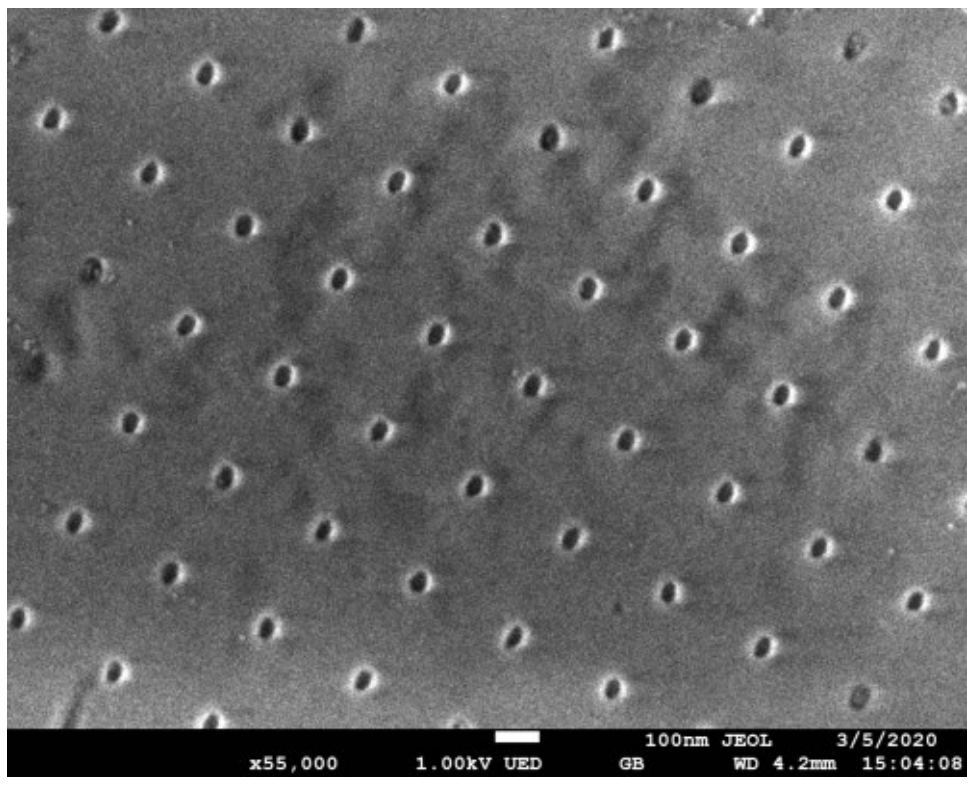
100nm JEOL 3/5/2020
x55,000 1.00kV UED GB WD 4.2mm 15:04:08

(a)

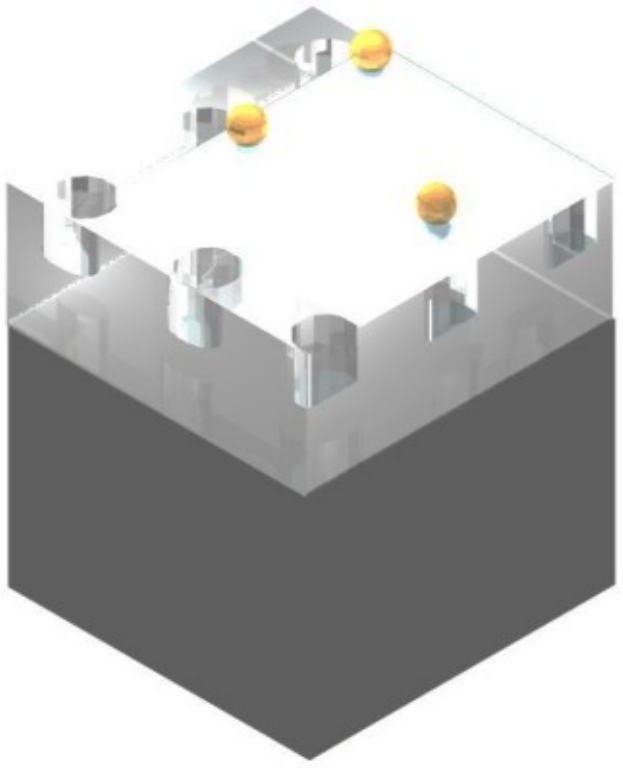

(b)

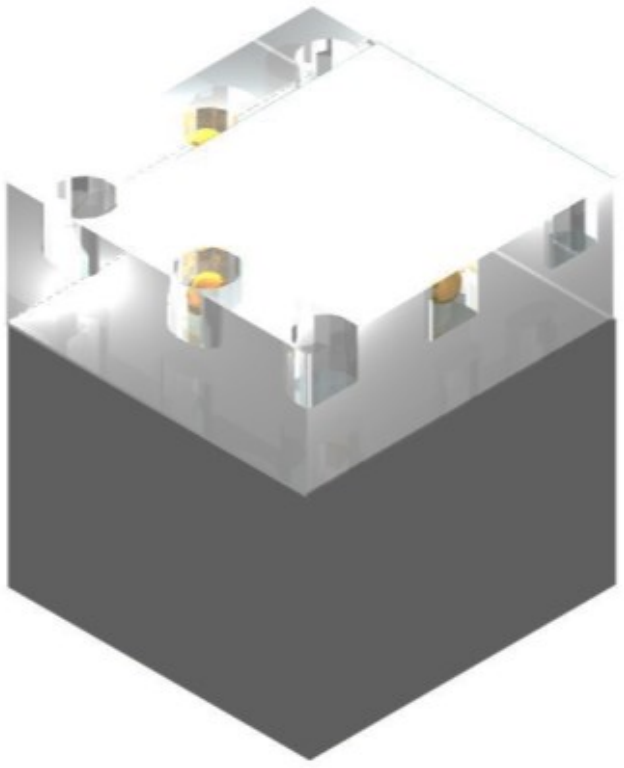

(c)

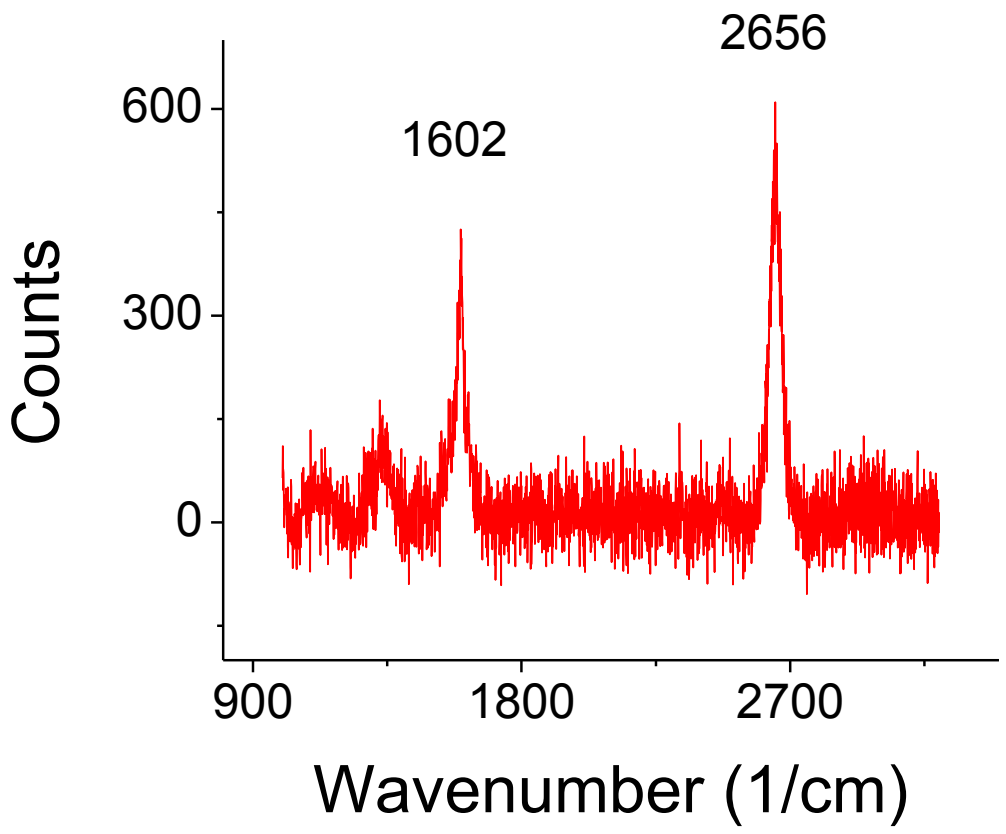

(d)

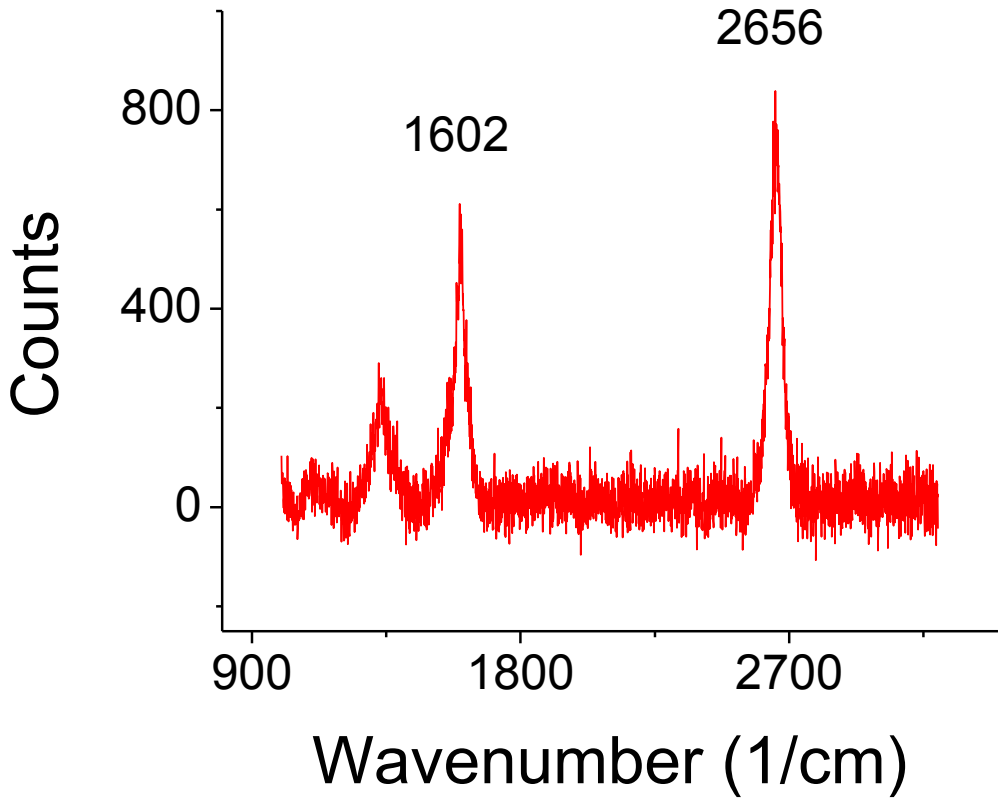

(e)

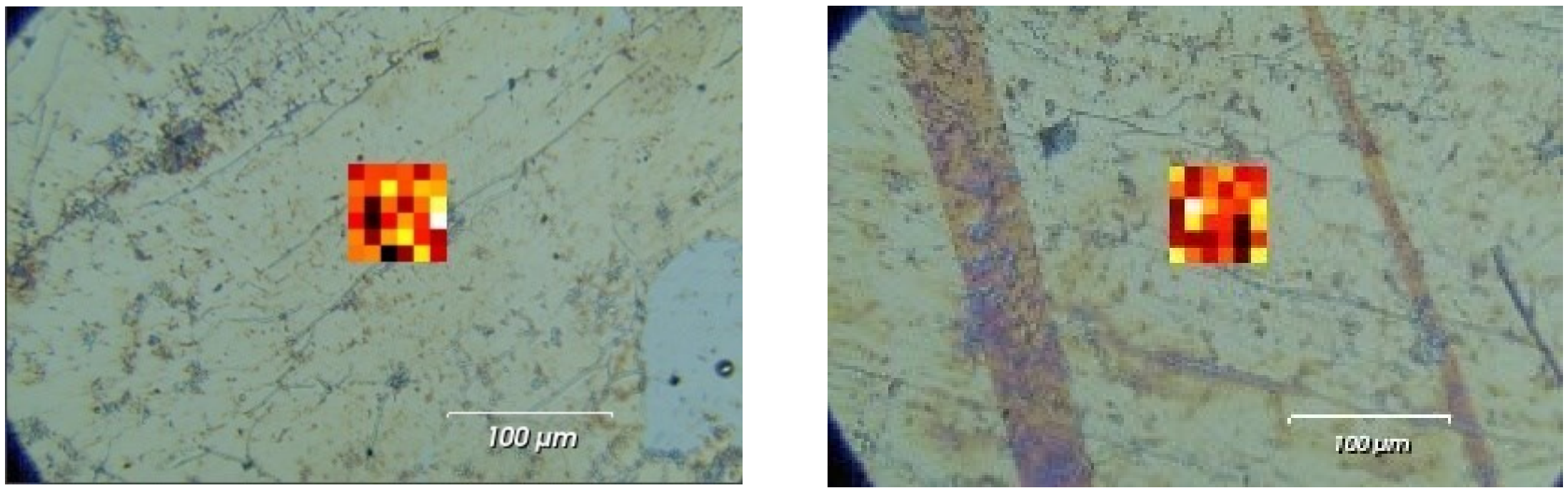

(f) (g)

Fig. 1. (a) SEM picture of the bare substrate with a hole array. (b,c) Schematics of samples S9 (spun QD on top of the graphene/hafnia surface guide) and S8 (spun and wiped QD under the graphene/hafnia guide). (c) Typical Raman spectra taken with a 2 mW, 633 nm HeNe laser; (d) when the dots were deposited on top of the hafnia/graphene layer while (e) is when the dots are deposited under the graphene. (f,g) Raman maps of the graphene's 2D line for (f) top and (g) under deposited dots allude to the monolayer nature of the graphene. The 2D intensity values, $I_{2D}$, for the black, red and white squares were, 500, 1000 and 3000, respectively. The ratio of $I_{2D}$ to the intensity of the G line, $I_G$ was approximately $I_{2D}/I_G$=1.3 throughout the scan.

A typical full fluorescence (FL) curve for sample S7 (with 10-hafnia layer on top of the graphene) exhibits a 470 nm line that is attributed to the 10-nm hafnia on the graphene (Fig. 2a). The line is missing from sample S2 that lacks the hafnia layer (Fig 2b), yet with a 250-nm thick PMMA on top of the graphene. The time-resolved curves, shown below, were obtained with a bandpass filter between 500 nm and 700 nm.

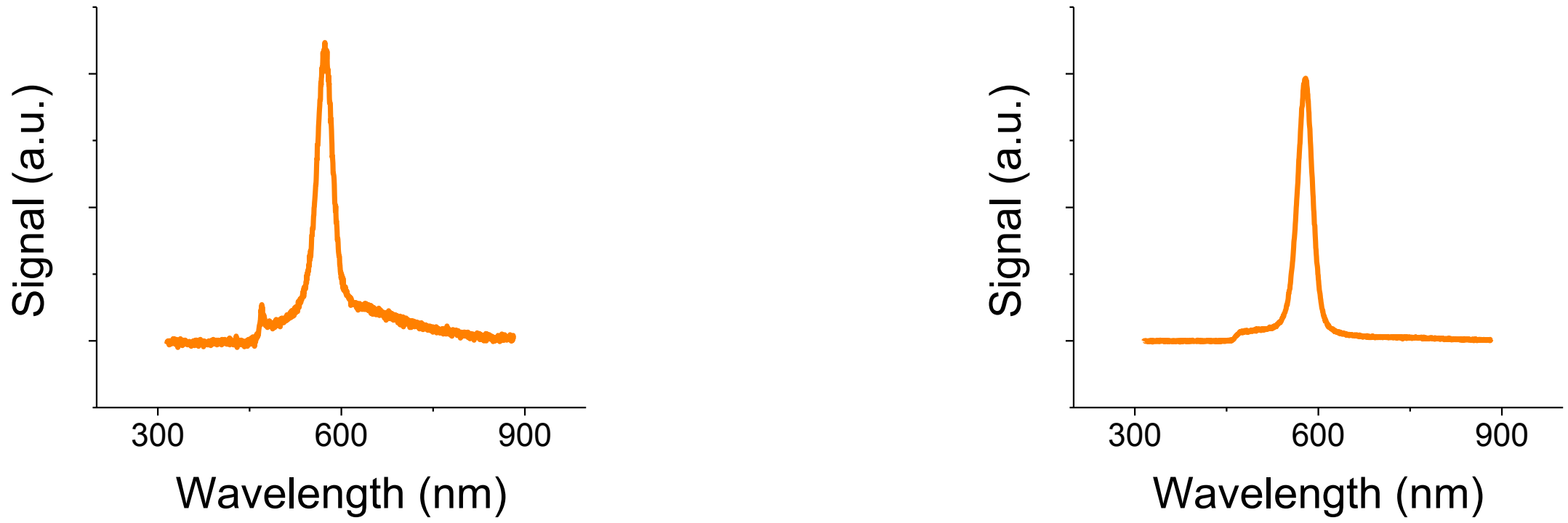

(a) (b)

Fig. 2. Full fluorescence curves for (a) sample S7 with 10-nm hafnia and (b) sample S2 without it. A laser cutoff filter was placed at 450 nm.

The effect of focusing the laser beam on the measured time rates is shown in Fig. 3 for sample S9 where the QD were spun over the hafnia/graphene layer. Three curves are shown for which the focal point was successively receding away from the sample surface. The peak intensity of the curve substantially varied for these three cases. This could be the result of: (a) the laser interrogated QDs that are at various distances from the quenching graphene layer, or that (b) the ach focal point interrogated different QD ensembles. The FL curves exhibited multiple decay rates and could be properly fitted with three decay constants. The largest decay rate is of the order of 2 $ns^{-1}$. It is attributed to dots that are at close proximity to the graphene guide. The medium rate is of the order of 0.2 $ns^{-1}$ and is attributed to dots that are less impacted by the graphene layer. The smallest decay rate is of the order of 0.05 $ns^{-1}$ and serves as a background component and could be also attributed to the photon life-time in the waveguide. As the focal point moves away from the graphene surface, the weight of the three ensembles is shifted towards dots that are less impacted by the graphene surface.

Here are some considerations to the fit process that led to the evaluation of the emission rates. (1) The geometrical effect of the laser spot on the overall emission rate assessment is not straight forward. The Gaussian beam has features of a plane wave only at the focal point, yet, excitations of the dots with varying degree of efficiency occurs with unfocused beam, as well.

(2) Limiting the fit to mainly one time component that is prevalent in a finite time range (namely limiting the fit, say, to a window of 10 ns after excitation) runs the risk that the solution will be affected by the boundaries of the time window.

(3) Having too many time constants may blur the physics of the processes.

(4) Considering the fit quality by only its R-square value is insufficient. One needs to consider the distribution of the residuals about the fitting parameter (see SI section). The residual distribution has to be evenly spread above and below the mean.

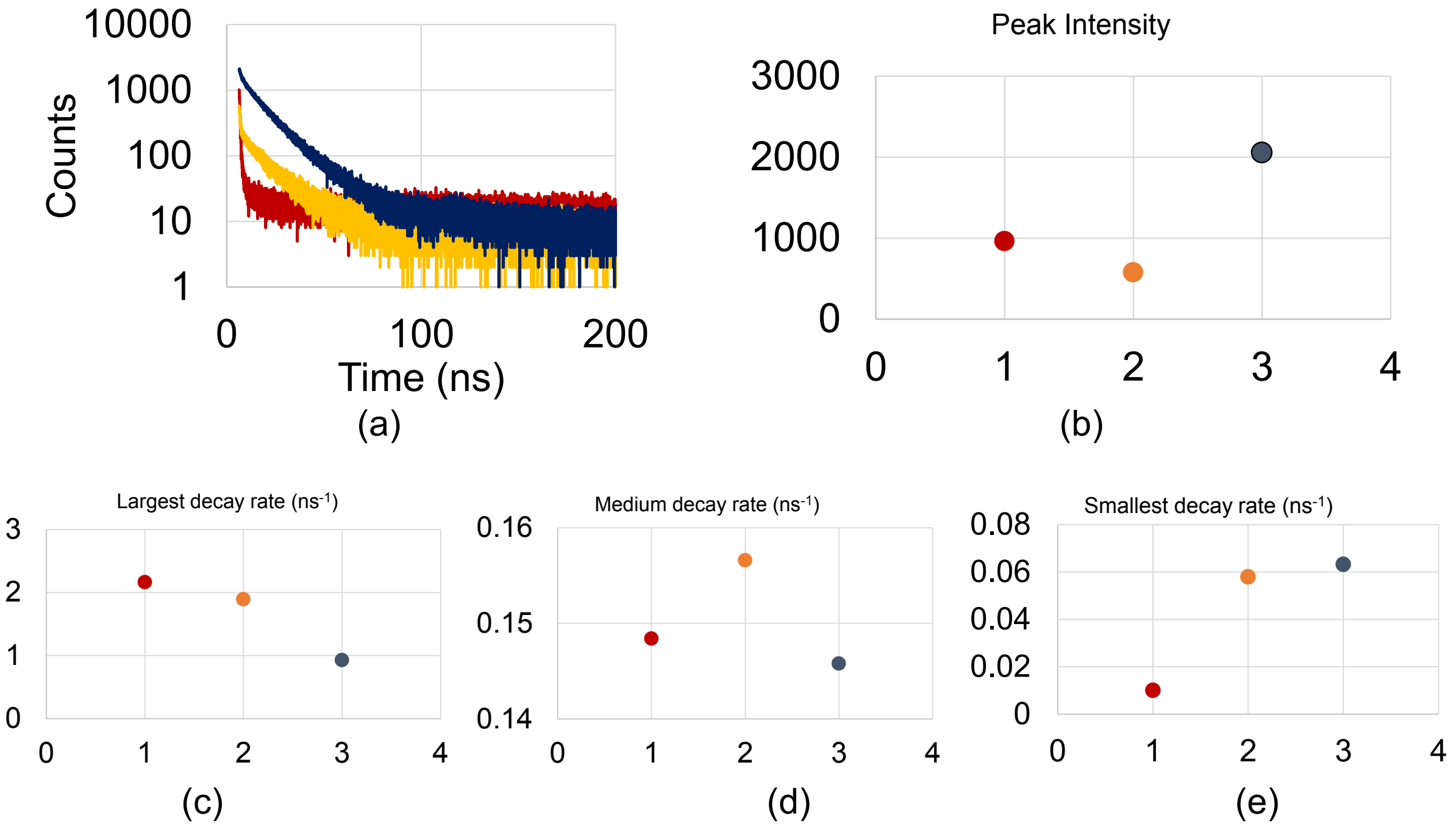

Fig. 3. (a,b) Red to blue: as the focus of the laser beam is receding away from the sample surface, the larger decay rate becomes smaller (c), the medium decay rate remains fairly constant and (d) the smallest decay rate increases.

A more detailed description is provided below for sample S7. The QDs were deposited on the hafnia/graphene layer by use of dip-coating. Plotting the maximum count vs azimuthal angle, $\phi$, between the laser polarization and the hole-array orientation yields mainly two peaks; at 0° and at 180° that allude to the stability and repeatability of the measurements (Fig. 4) but do not clearly exhibit much symmetry that can be related to the square hole-array.

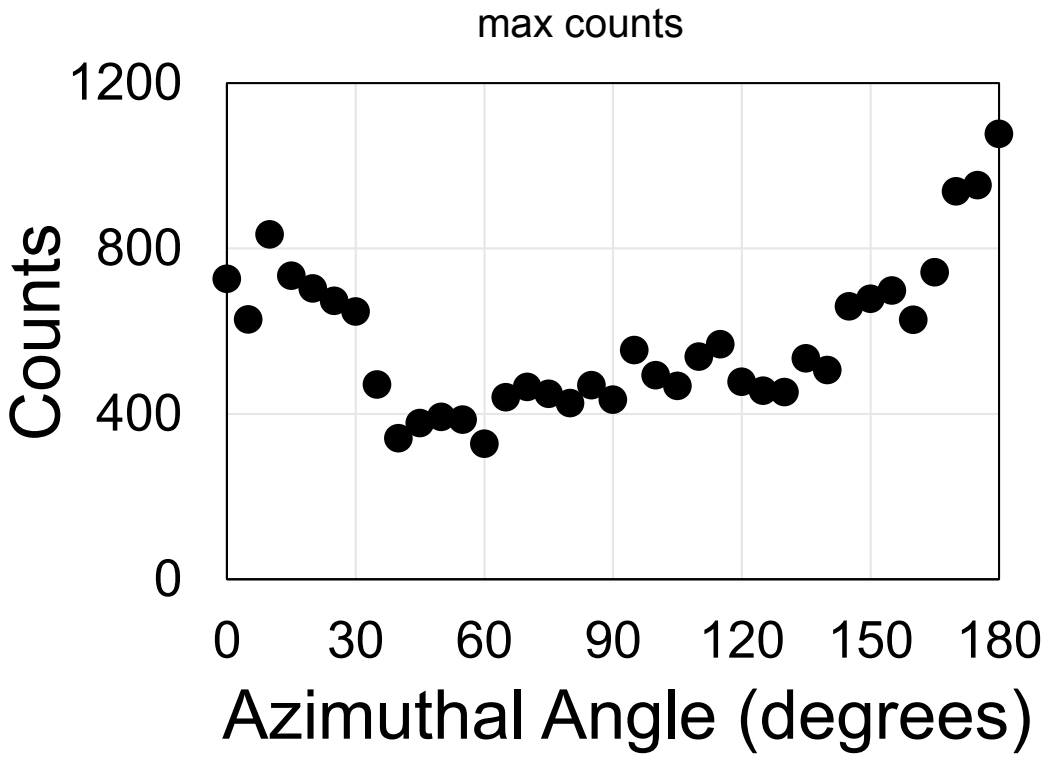

Fig. 4. Variations of the peak intensity as a function of azimuthal angle, $\phi$ between the laser polarization and the hole-array orientation.

We concentrate on the first two major decay rates: the largest (of ca 2 $ns^{-1}$) and the next smaller (of ca 0.2 $ns^{-1}$) ones. The smaller decay rate distribution as a function of the azimuthal rotation angle $\phi$ is shown in Fig. 5a. The larger decay rate is shown in Fig. 5b. The smaller decay rate is equivalent to an average life time constant of 6 ns. That value is within an order of magnitude for a stand-alone QD (on a 10 ns scale). The larger decay rate is equivalent to a life-time constant of 0.5 ns. It exhibits more pronounced 90° cycle as indicated by the blue line and as expected by the square nano-hole symmetry. The blue line also indicates that the coupling constant, $\kappa$, and the interaction length between surface mode and the hole-array planes behave as $\kappa L \sim 1$.

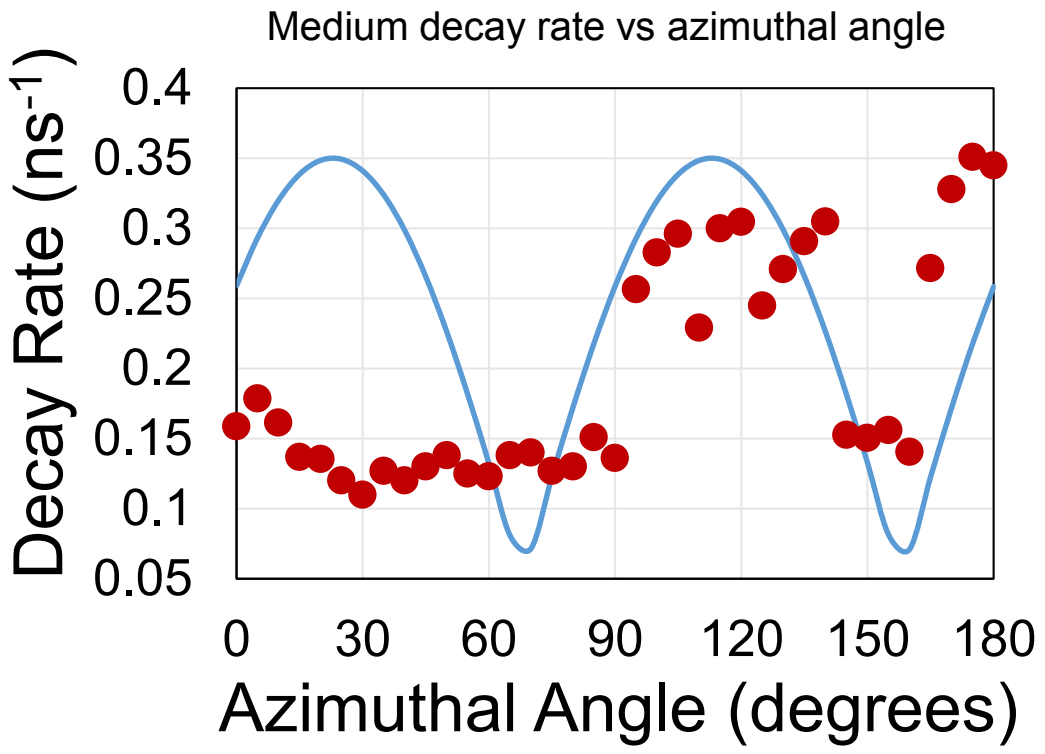

(a)

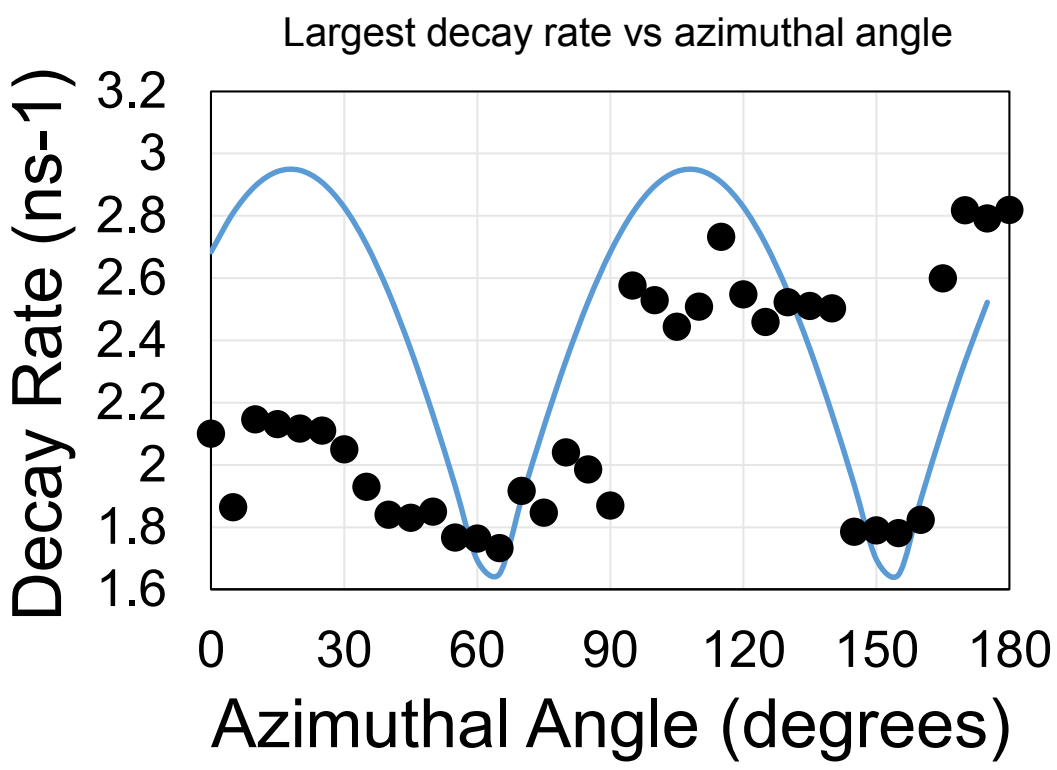

(b)

Fig. 5. Sample S7. There are essentially two decay rate coefficients: (a) below and (b) above 1 $ns^{-1}$. The blue line in (b) is guide to the eyes (see SI section). The curve was shifted by $\phi_0$ since the initial hole-array orientation was unknown. The error in the decay rate fit was less than 1% (and hence is contained within a data point; the error in the azimuthal angle was 0.5°. The error in repeating the measurement of the same spot is less than 10%. In general, QD which are deposited on top of the graphene exhibit clearer undulations.

The error in the decay rate fit is less than 1% (and hence is contained within a data point; the error in the azimuthal angle is 0.5°. We attempted to maintain the same spot position during sample rotation. The error of maintaining that spot is estimated at less than 10%. Yet, uncertainties in the exact spot position may have contributed to coefficient variations.

The larger decay rate coefficient as a function of azimuthal angle $\phi$ for samples S8 (QD under the graphene) and S9 (QD on top of graphene) are shown in Fig. 6a,b (see also SI section). Unlike sample S7, here the QD were spun over the surface and their concentration was less than S7 (25% of S7 concentration). The data undulations are more pronounced for QD placed on top of the graphene, yet 'cleaner' for QD deposited underneath it. Similar emission rates for top coated and under--coated QD are the result of similar distance from the graphene layer. The 90° symmetry is consistent with $n_{eff}$=1.15. Fig. 7c shows data for sample S9 when keeping the same spot position and maximizing the FL intensity (as opposed to focusing the spot onto the sample surface). The range of the larger decay rate is similar to the overall range of Fig. 6b, yet without an apparent undulations.

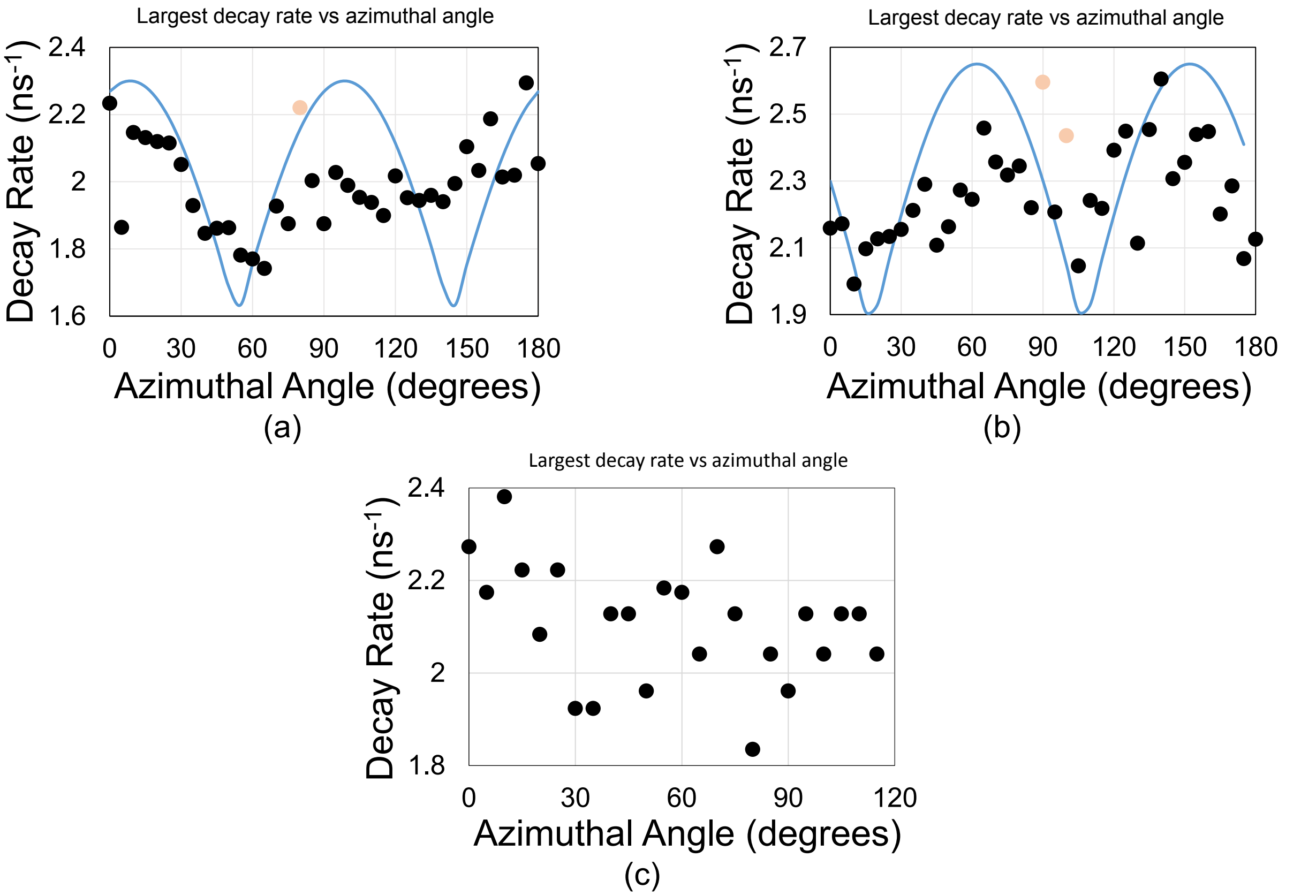

Fig. 6. (a) Sample S8 (QD in the holes under the graphene guide) and (b) sample S9 (QD on top of the 10-nm hafnia on the graphene guide). The QD were spun over the samples. The blue line is but a guide for the eyes. The red dots indicate outliers: a solution was achieved with a good R square value of above 0.97 but residuals were not evenly distributed about the mean (see SI section). (c) Maximizing the fluorescence signal (as opposed to focusing onto the sample surface) resulted in decay rates that covers similar value range to (b) but failed to uncover meaningful undulations. The blue line is the expected undulations.

If surface guide is interfaced with a relatively top thick polymer instead of air the surface guide becomes more symmetric. The 250-nm PMMA layer, used during the transfer stage of graphene was retained and no oxide was deposited on top of the graphene. Sample S2 was made of spun QD in the nano-hole array and under the graphene layer. Judged by the emission rates, the dots resides away from the graphene. Unlike the previous samples, both the longer and smaller emission rate coefficients exhibit 45° undulations. This was made with the (½,0) planes, or every other plane. The Bragg peaks appear every 45° and the reflectivity is much narrower than for a air-top guide. (SI section)

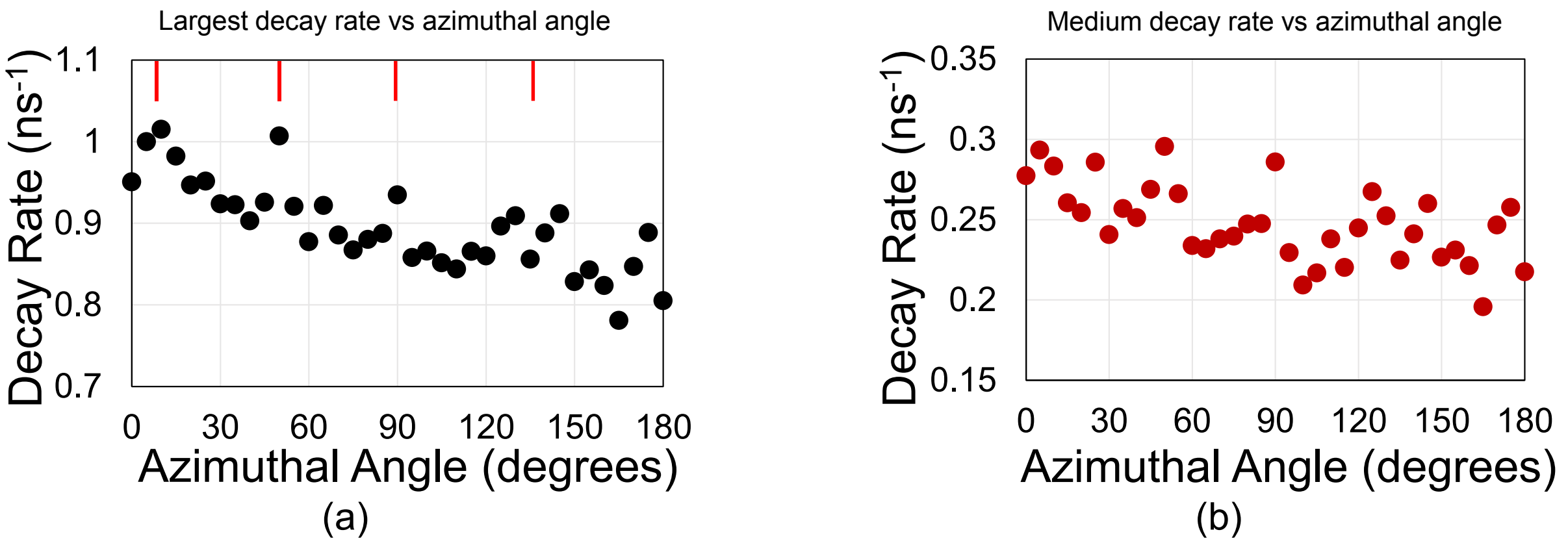

Fig. 7. Decay rate coefficients in $ns^{-1}$ for sample S2 exhibiting 45° symmetry for both, (a) the largest rate and (b) the medium rate. The trend for the latter is not as clear as for the larger rate. The red lines point to the peaks.

## Conclusions

We observed variations in the quenched fluorescence's life-time of QD embedded with patterned quasi two-dimensional graphene surface guides upon azimuthal rotations. These variations were as large as 50%. Since coupling to spatially resonating surface modes is also associated with large emission rates for nearby chromophore, one in principle, could control an energy transfer from one type of dots to another via the graphene surface guide. If properly designed, spatial perturbation may not only control the chromophore emission rate but also enable an efficient fluorescence detection at particular directions.

## Methods and Experiments:

All samples were made on a 500-microns thick p-Si wafers coated with 150-nm $SiO_2$. A square hole-array, with a pitch of 250-nm, a hole-diameter of 30-nm and a hole-depth of ca 30-nm was defined by e-beam lithography and etched into the $SiO_2$ layer. A monolayer graphene was deposited over the hole-array. The graphene was coated with 10-nm hafnia by atomic layer deposition (ALD) prior to the graphene transfer as per our recipe described elsewhere [24]. Note that the ALD is made at a relatively high-temperature of ca 200 °C, which may prohibit its use when the QD are already situated in the hole-array. The QDs (core/shell, CdSe/ZnS [27]) were purchased from Mesolight and were deposited either on top of the hafnia/graphene layer (Fig. 1a) or underneath it within the holes (Fig. 1b). For the latter case, special attention was given to maintain as many dots inside the holes and remove excess dots from the oxide surface (where direct contact is made with the graphene). When the QD are embedded in the holes, the filled holes may accommodate only one dot per hole since the dot is coated with a ligand whose overall diameter is ca 20-nm [13]. The dots are situated at the hole's bottom and the separation between the dot and the graphene would be 20 nm (the hole's depth minus the radius of the ligand coated dot). When the dots are deposited on the top of the 10-nm hafnia, the separation between the dot and the graphene can be more accurately maintained, being 20-nm, as well.

Raman spectra were taken with a 633 nm HeNe laser at an intensity of 2 mW as an excitation source and a x20 objective. Stresses in the graphene from the hafnia and the QDs may affect its 2-D line, albeit its position remained in the vicinity of 2650 1/cm. The QD emitting in the wavelength of ca 575 nm were pumped with a 90-ps, 250 $\mu$W 405 nm laser at a pulse rate of 25 MHz. The fluence was 1000 W/cm$^2$. The dots, suspended in toluene were either dip-coated or spun at 2500 RPM for 30 s.

For the time-resolved and fluoresce measurements, the laser beam at 405 nm was focused by an x5 objective to a ca 25 micron$^2$ spot onto a well-defined spot that was visible through the set of filters and could be visited time and again. The sample area could be viewed using an optical microscope that could be separated from the measurement system by a prism and which was equipped with a white light illuminator and a CCD camera while viewing through the same objective. Error in repeating the measurement of the same spot was estimated as less than 10%. Uncertainties in the exact spot position may have contributed to coefficient variations. For the fluorescence data, the detector was equipped with a cut-off low-pass filter whose cut-off wavelength was 450 nm. For the time-resolved measurements, a bandpass filter between 500 nm and 700 nm was used (with a different detector than the one used for the fluorescence measurements).

**Acknowledgement**

This work was performed, in part, at the Center for Nanoscale Materials, a U.S. Department of Energy Office of Science User Facility, and supported by the U.S. Department of Energy, Office of Science, under Contract No. DE-AC02-06CH11357. We thank Prof. D. Ko from NJIT for availing his dipping system and his insightful comments.

The effect of periodic spatial perturbations on the emission rates of quantum dots near graphene platforms and related energy transfer topics

X. Miao[1], D. J. Gosztola[2], X. Ma[2], D. Czaplewski[2], L. Stan[2] and H. Grebel[1]*

[1] Electronic Imaging Center and ECE Dept., New Jersey Institute of technology (NJIT), Newark, NJ 07102, USA. grebel@njit.edu
[2] Center for Nanoscale Materials, Nanoscience and Technology Division, Argonne National Laboratory Argonne, IL 60439

Supplementary Information

**Rate Equations:**

The system is pictorially presented in Fig S1:

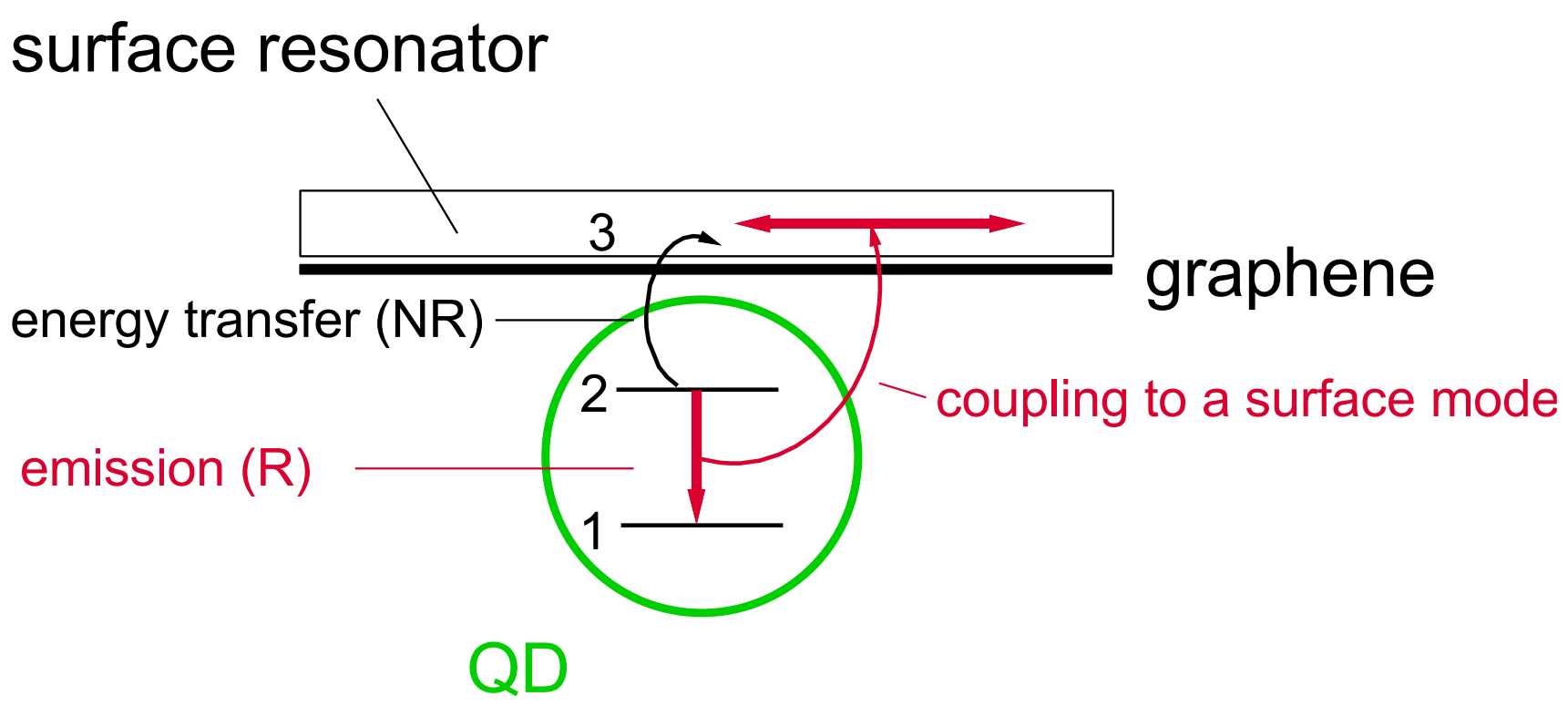

Fig. S1. Schematics of the QD coupled to graphene and a surface resonator.

We assume a three-level system. A pulse R(t<0) excites the QD to level 2 from level 1. The emission may be coupled non-radiatively (NR) to level 3 at time $\tau_{23}$; it may emit a photon at time $\tau_{21}$ and couple radiatively (R) to the resonating surface mode. The radiative emission rate, $\Gamma_{2\rightarrow1}=1/\tau_{21}\sim\rho(\nu_{21})H_{21}$ includes the interaction term $H_{21}$ and the density of the final state, $\rho(\nu_{21})$ [Fermi's golden rule, S1]. The density of states (DOS) for a resonating mode is basically 1 mode per mode's width $\Delta\nu$ per volume V, or, $1/(\Delta\nu * V)$. The mode spectral width is written as: $\Delta\nu^{-1}=Q/\nu_{21}$, with Q - the resonator's quality factor and $\nu_{21}$ - the transition frequency; the mode volume is: $V\sim(p*\lambda/2)^2$, where p – the characteristic decay length away from the surface and $\lambda_{21}=(c/n)/\nu_{21}$ – the mode wavelength. The characteristic decay length of the mode, p, is a fraction of a wavelength, typically, $\lambda/4$ near conductive surfaces. Thus, for a surface resonator, the density of states is increased by the quality factor Q compared to a free-space QD and consequently, the spontaneous emission rate is increased by Q, as well [Purcell's effect, S2, S3]. We assume that when the transition frequency coincides with the resonator mode, the transition rate is dominated by the DOS of the resonator.

We use the following rate equations for a pulse pump, R:

$dn_2/dt=-n_2/\tau_{21}-n_2/\tau_{23}-\sigma n_2 n_{ph}+R(t<0)$ (S1)

$dn_3/dt=-n_3/\tau_{33}+n_2/\tau_{23}+\sigma n_3 n_{ph}$ (S2)

$dn_{ph}/dt=-n_{ph}/\tau_{ph}+\sigma n_2 n_{ph}-\sigma n_3 n_{ph}$ (S3)

Here: $n_2$ is the excited electron density, $n_3$ is the excess electron density in the graphene due to non-radiative energy transfer, $n_{ph}$ is the output photon density, $\tau_{21}$ is the life-time of the radiative emission (coupled to the surface resonator), $\tau_{23}$ is the non-radiative transition to the graphene, $\tau_{33}$ is the dissipation time and $\tau_{ph}$ is the photon life-time in the surface resonator. The cross-section, $\sigma$ is assumed to be equal for the lossy photons; $\tau_{ph}$ is rather short.

The solution of Eq. 1, is $n_2(t)=n_2(0)\exp(-t/\tau_{2eff})$, with $1/\tau_{2eff}=1/\tau_{21}+1/\tau_{23}+\sigma n_{ph}$; the stand-alone radiative emission rate of the QD,$1/\tau_{21}$ is increased by the non-radiative transfer of energy to the graphene and the presence of increase mode density in the resonator.

Let us assume that the presence of $n_3$ is only due to $n_2$, $n_3=\alpha n_2$.

The excess charge in the graphene is, $n_3(t)=n_3(0)\exp(-t/\tau_{3eff})$, with $1/\tau_{3eff}=1/\tau_{33}-1/\alpha\tau_{23}-\sigma n_{ph}$. Interestingly, the rate of exchange energy (negative sign for a gain), $-1/\alpha\tau_{23}$ is accentuated by the weak coupling between the QD and graphene ($\alpha<<1$). Weak coupling alludes to larger distances between the QD and graphene and the interaction term $H_{23}$ that enables the coupling substantially decreases as a function of distance [S4].

**Numerical Assessments:**

The coupling to surface modes is pictorially shown in Fig. S2.

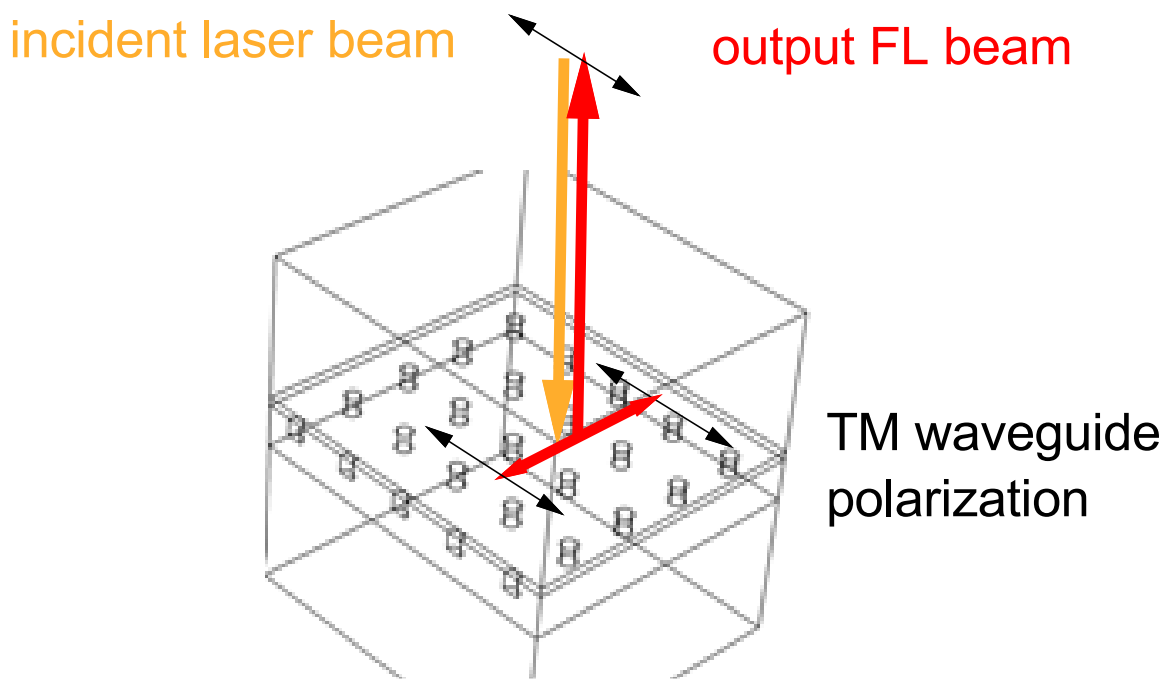

Fig. S2. An s-polarized incident beam forming two counter propagating TE waveguide modes. The polarization (black arrows) is preserved due to fast non-radiative life-time. The waveguide supports both TE and TM modes.

The various electric field distributions are evaluated at a cut-plane at the interface between the waveguide and the substrate (Fig. S3). The model was constructed with a CAD tool (COMSOL). Scattering boundaries were used around the structure. These are equivalent to perfectly matched layers (PML) to avoid back reflections. A thick Si wafer (bottom) is covered by a 150 nm silica film, which is decorated with air pillars of depth 50 nm. The pillars of radius 30 nm are covered with a surface guide and are topped either by air, or a polymer film with refractive index similar to that of the silica. The pitch of the hole-array is 250 nm. The surface waveguide is composed of graphene, 10-nm hafnia and QDs. The QD (with a radius of ca 3-4 nm) and their ligand coating have an effective thickness of 20 nm. The effective thickness of the guide is 30-nm and its refractive index is 2.4+i0.24. As we shall see below, the actual refractive index of the optical surface guide is of little consequence because most of the mode intensity propagates outside it. In Fig. S3 we show two cases: (1) a collection of many dots that form a plane wave at the emission wavelength of 575 nm along the x-direction and polarization along the y-direction (the waveguide's TE mode) and (2) an emission from a single trapped dot as a spherical wave. The electric fields, polarized along the y- and z-directions were assessed.

(1) A plane wave of 1 V/m and whose polarization is along the y-direction (parallel to the guide surface) propagates along the x-y plane of the surface guide. Fig. S3a shows the $E_y$ component (parallel to the guide surface) and Fig. S3b is for Ez polarization (perpendicular to the guide surface). The effective index of the surface guide can be deduced when referencing the wavelength along the interface to the array pitch of 250 nm. Thus, $\lambda_n=\lambda_0/n_{eff}$~2a for air topped sample fulfilling the Bragg condition along the x- and y- directions. This is translated to $n_{eff}$~1.15 which is consistent with the experiments. Similarly, for a guide surrounded by polymer and silica, $n_{eff}$~1.45. Both cases allude to the fact that the wave travels mostly outside the surface waveguide. Interestingly, if the average permittivity of the air/quartz interface, $n^2_{eff}=\varepsilon_{eff}=(\varepsilon_{air}+\varepsilon_{silica})/2=(n^2_{air}+n^2_{silica})/2$, or $n_{eff}=1.25$.

(2) The case where the waveguide is excited by a spherical point source (QD), which is situated in one of the holes is presented in Fig. S3c,d. The wavelength match is $\lambda_0/(n_{eff}=1.15)$~2a and a higher orders, for air topped and polymer topped, respectively. Another intuitive view is to consider the graphene/hafnia/OD interface as either asymmetric guide when the bottom layer is made of silica and the top layer is air, or, a symmetric guide when the top layer is made of a polymer. In either case, most of the surface mode is propagating outside the guide and simulations imply that the propagation is above waveguide cut-off. Finally, Fig. S3e shows the Ez polarization component (perpendicular to the guide surface). The component is not zero and is concentrated in the pillars. Thus, a y-polarized spherical wave excites a z-componentwhich are mostly concentrated in the hole-pillars.

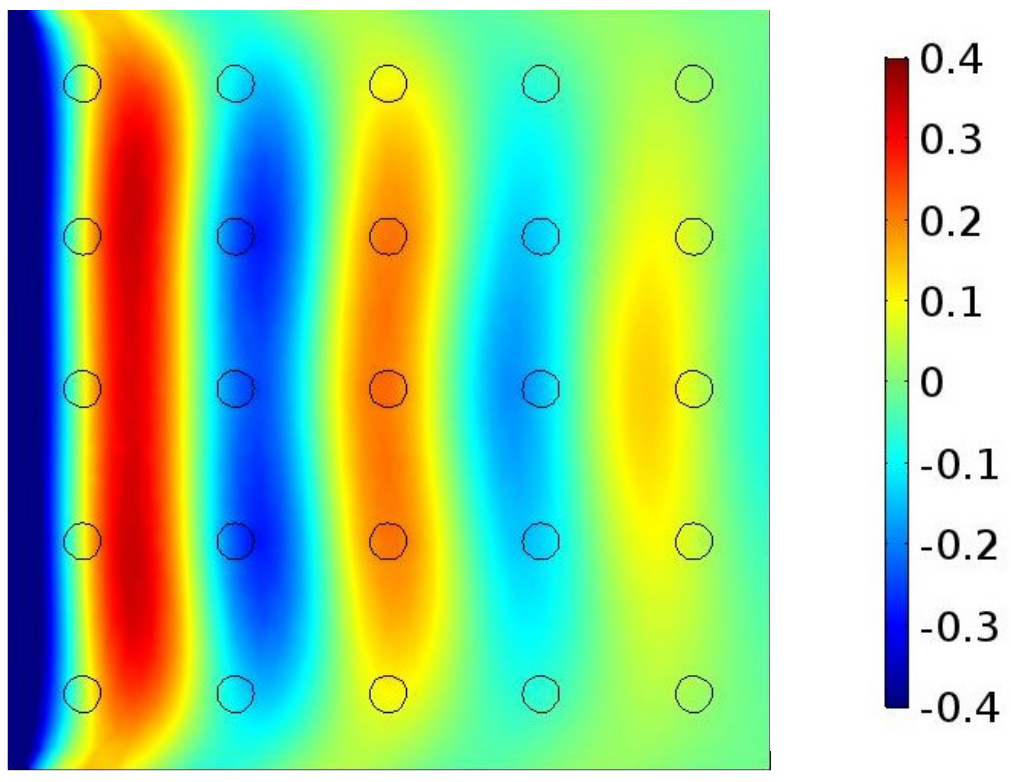

(a)

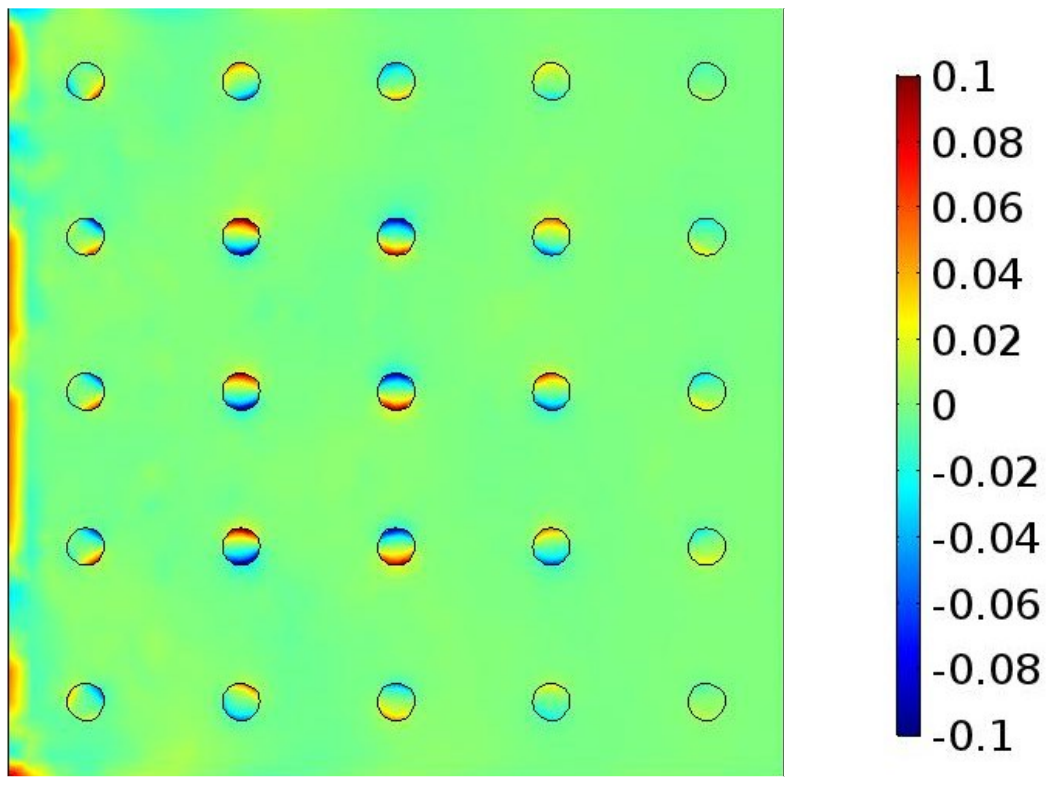

(b)

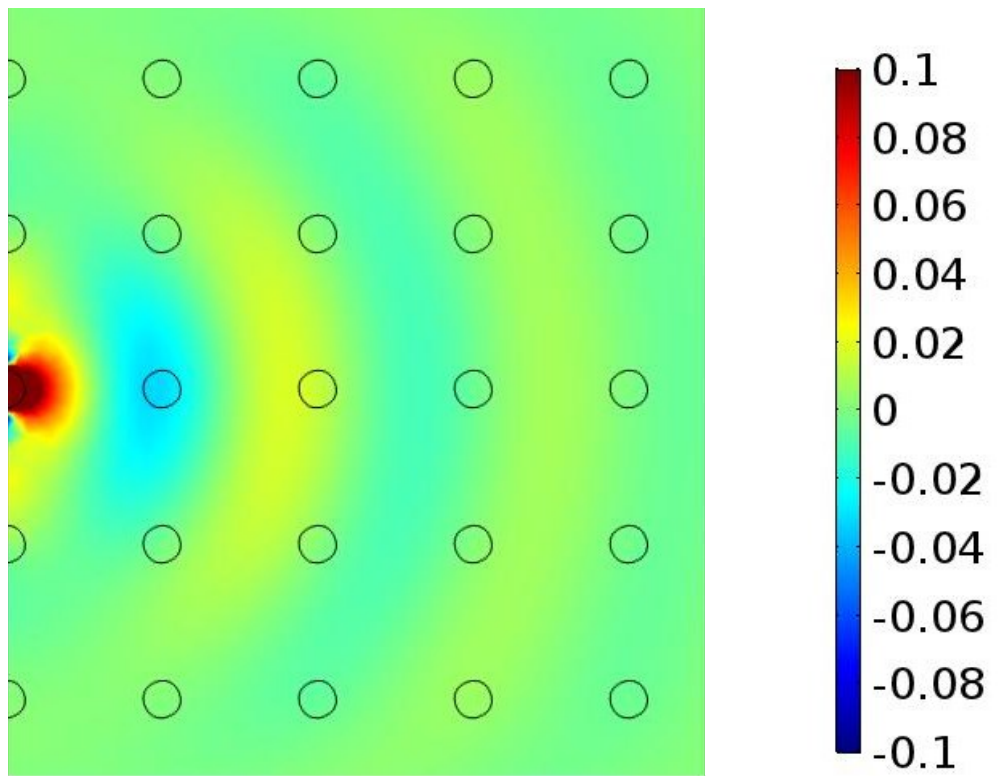

(c)

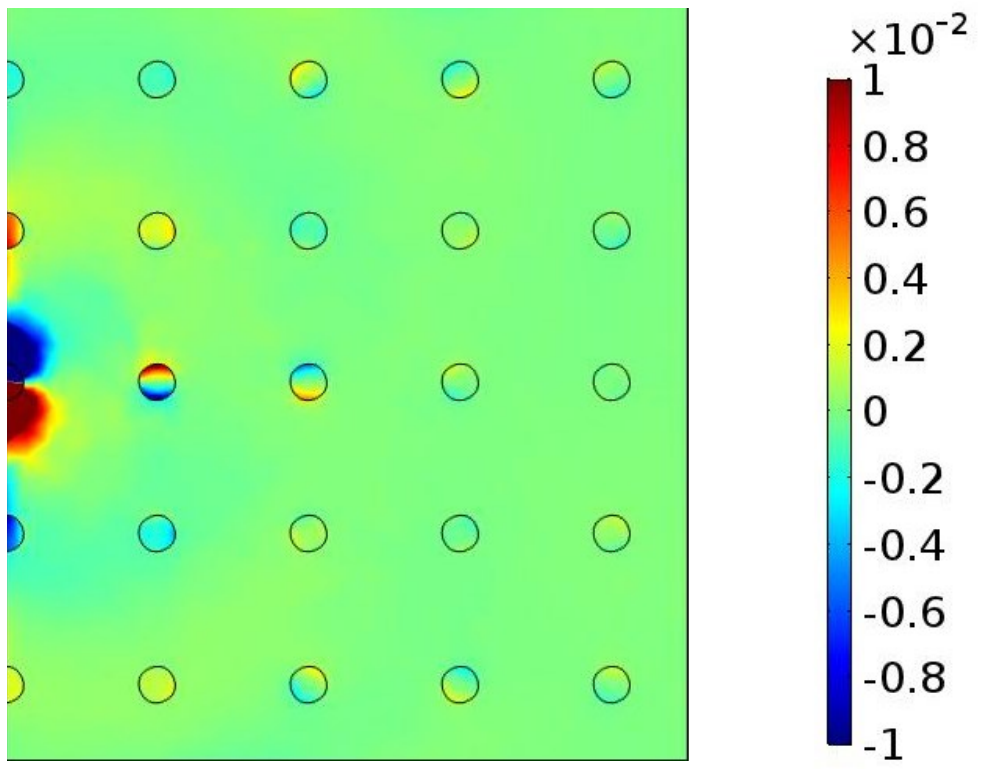

(d)

Fig. S3. (a) A plane wave is excited from the left (x=0) and let to propagate in all directions. Portion of the wave is captured by the surface guide. Shown is $E_y$ (parallel to the guide surface) at the effective surface guide between guide and quartz. Note the focusing of the beam by the sub-wavelength hole-array. (b) $E_z$ (perpendicular to the guide surface) at the effective surface guide between guide and quartz. (c) $E_y$ for a spherical wave excited by a QD in one of the holes on the left and is let to propagate along all directions. The surface guide is made of air-guide-quartz layers. (d) $E_z$ at guide-quartz interface for a spherical wave for a polymer-guide-quartz layers. The emission wavelength was 575 nm and the intensity legends are in V/m.

At normal incidence, we may pick up the x- and y- along the square hole-array coordinates. **G** is the reciprocal wave vector of the spatial square array of holes with a pitch $\Lambda$; $G_x=G_y=G$. Coupling to and from the hole-array at normal incident fulfil, $\beta_s=G[q_2^2+q_2^2]^{1/2}$ with $q_{1,2}$ – integers, and $\beta_s=2\pi\lambda/n_{eff}$ – the wavenumber of the surface mode. The Bragg condition is $|\boldsymbol{\beta}_s-\mathbf{G}|=\beta_s$. Thus, $2\Lambda\cos\phi=m\lambda/n_{eff}$. At normal incidence, we may pick up the x- and y- along the square hole-array Once coupled to the surface mode, we can approximate the in-plane reflections as (counter propagating coupled mode theory [S5],

$$R=\frac{\kappa^2 sh^2(s\cdot L)}{s^2 ch^2(s\cdot L)+(\frac{\Delta\beta_s}{2})^2 sh^2(s\cdot L)},$$

where, $s^2=\kappa^2-(\frac{\Delta\beta_s}{2})^2$, $\kappa$ is the coupling constant between the hole-array planes and the propagating guided mode, and $\Delta\beta_s=2k\cos(\phi)-qG$ and $L$ is the effective interaction length. Fig. S4 shows a super imposed curve for the Bragg scatterings from the x- and y-directions. For $\kappa L\sim1$, the curve can be simply approximated by $|\cos(2\phi)|$ (magenta curve) and was used to accentuate the azimuthal curves.

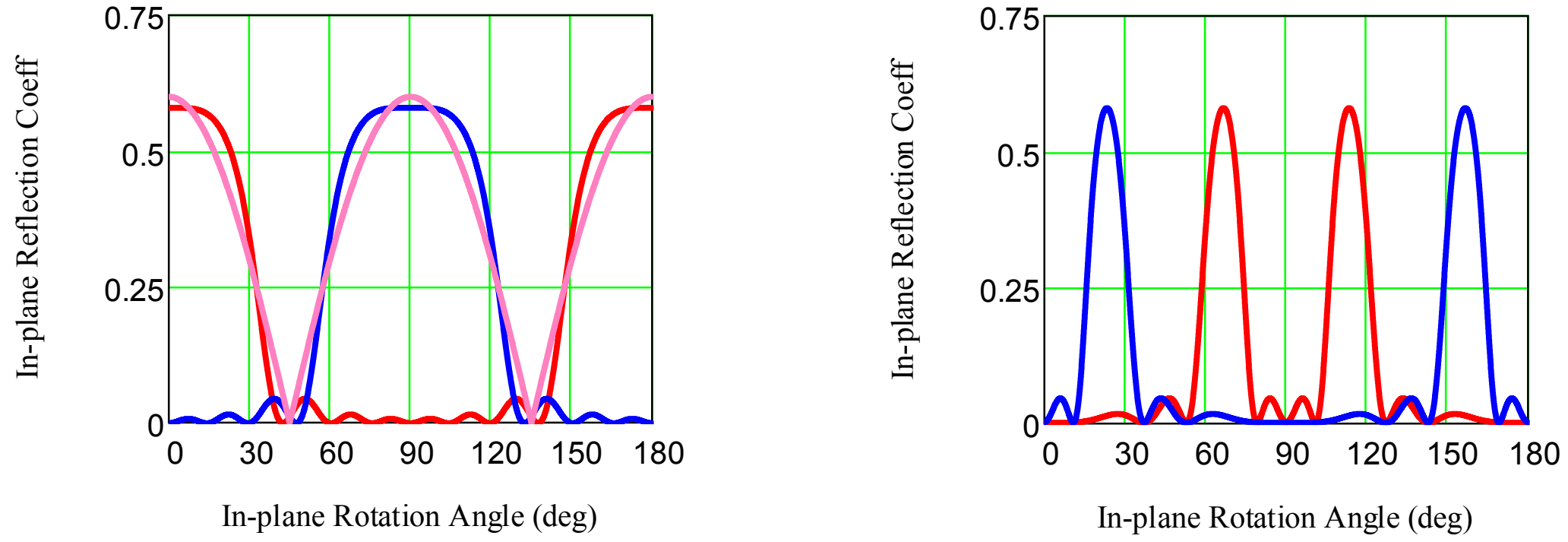

(a) (b)

Fig. S4. (a) Superimposed curves of Bragg scatterings (red and blue curves) corresponding to the scattering by the x- and y-planes of the hole-array for κL~1 when the surface guide is (a) topped by air and (b) topped by a polymer. For (a), the scatterings were made by the (10) planes. We used a simplified approximation ~│cos(2ϕ)│ (magenta) to accentuate the trend. For (b), the scatterings could be made by the (½,0) planes (as shown) for every other plane, or alternatively by (1,0) and (1,1) planes. The latter condition requires a much large coupling constant, though, κL~3. The peaks in (b) are clearly narrower, corroborating Fig. 8 in the text.

In Fig. S5(a,b) we provide curve fitting examples for reference points on the oxide without graphene for sample S9: (a) outside the hole-array region and (b) inside it. Time constants for the inside the hole-region have been reduced. The relevant rate (the larger decay rate) has been increased from ~1 $ns^{-1}$ outside the hole-region to1.6 $ns^{-1}$ inside it.

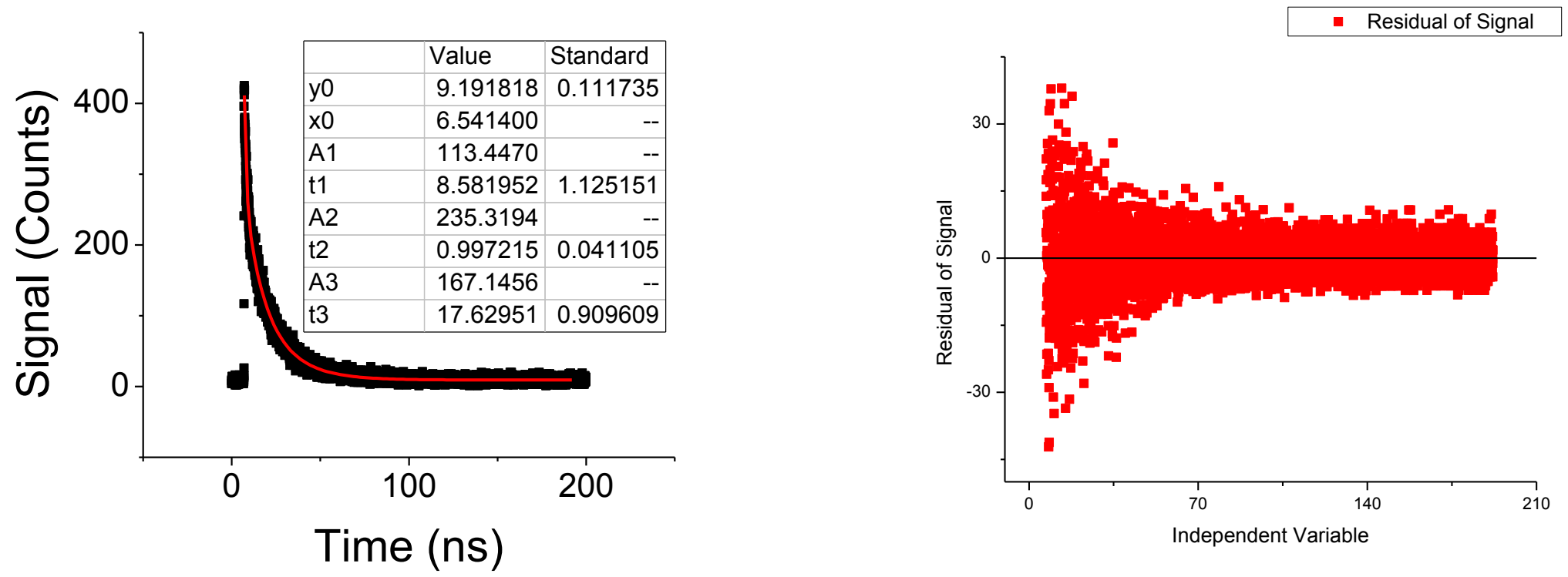

(a)

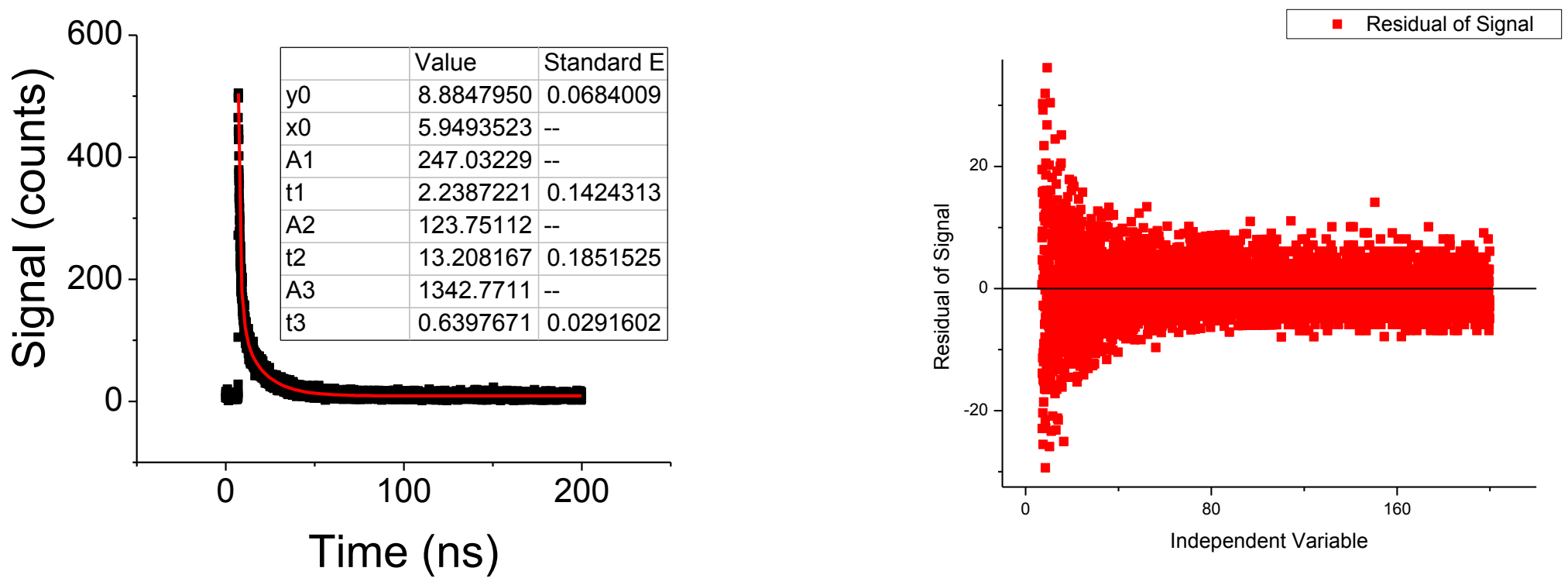

(b)

Fig. S5. Good fits with proper residuals (a,b) and improper residual distribution. (a) Bare oxide outside the hole-region. (b) Inside the hole-region – the time constants have been substantially reduced.

As discussed in the text, a good fit ought to consider not only its convergence but the distribution of its related residuals. As shown in Fig. S6, the point for sample S9 at $\phi$=100º is considered an outlier because the residuals are not evenly distributed above and below the zero-line (in other words, the data points are not completely random).

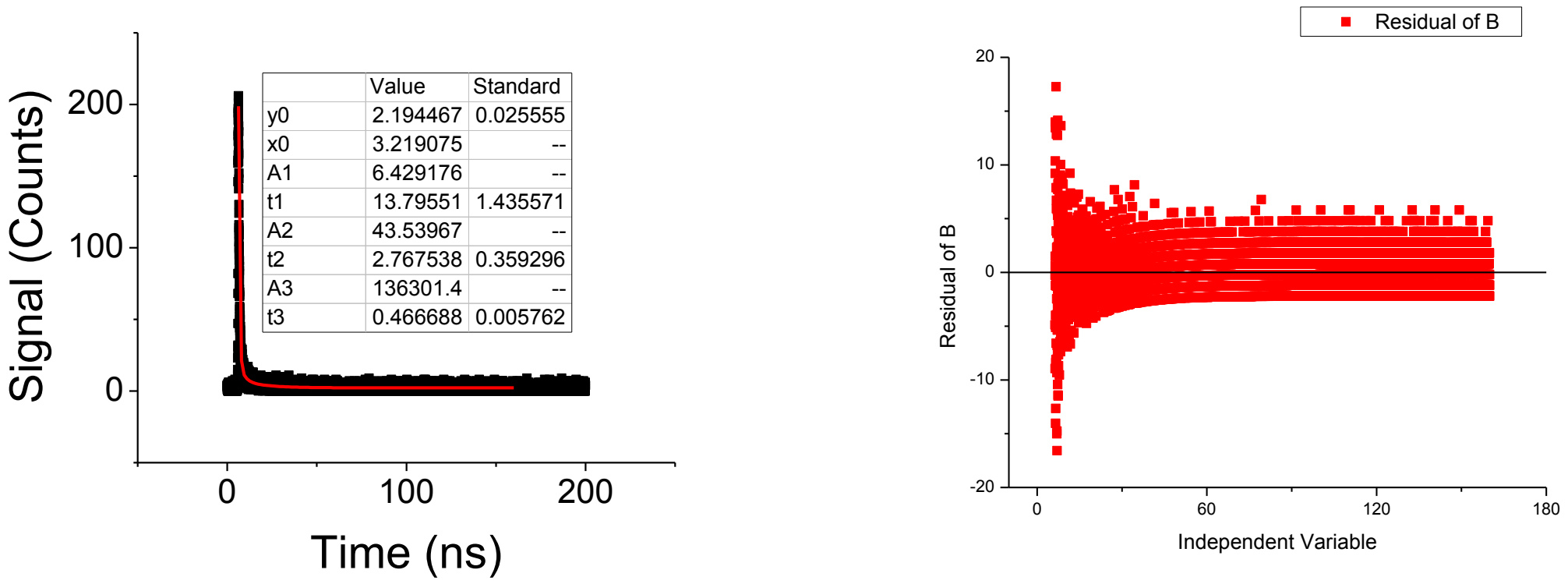

(c)

Fig. S6. (c) S9 outlier at $\phi$=100º. The R-square=0.95, yet the residuals are not distributed evenly and tilted towards the positive part. In addition, the simulated peak is shown to start at 3.2 ns, shifted from ca 6 ns for the experiment.

Figure S7 shows the 'medium' decay rate for samples S8 (QD below the graphene inside the holes) and for S9 on top of the hafnia above the graphene layer. The undulations are not as clear as the 'largest' decay rates.

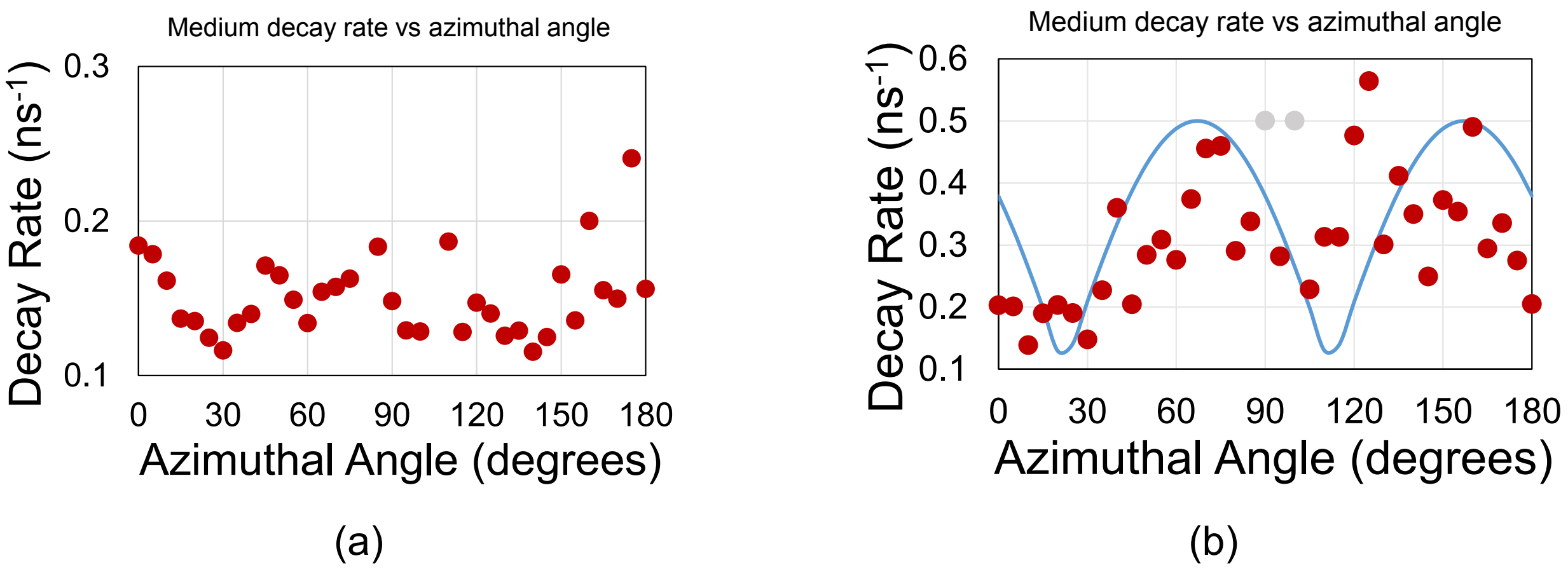

Fig. S7. (a) S8 (QD below the graphene inside the holes) and (b): for S9 on top of the hafnia above the graphene layer. The grey dots are outliers.